\documentclass[letterpaper, 10 pt, conference]{ieeeconf}  

\IEEEoverridecommandlockouts     
                                    
\usepackage[caption=false,font=footnotesize]{subfig}
\usepackage{graphicx} 
\usepackage{cite}
\usepackage{algorithm}
\usepackage{amsmath}
\usepackage{amsfonts}
\usepackage{algpseudocode}
\usepackage{booktabs}
\usepackage{comment}
\usepackage{url}
\usepackage{caption}
\usepackage{hyperref}
\usepackage{xcolor}

\title{Value-Aware Prediction for Robust Multi-Agent Coordination Under Communication Loss}

\author{Kemal Devrim Kafadar$^{1*}$,
        Eren Özaltun$^{2*}$,
        Mahmud Efnan Şanlı$^{3}$,
        Feyza Orak$^{3}$,
        Emirhan Gazi$^{3}$,
        \\Kubilay Kağan Kömürcü$^{3}$,
        Nazım Kemal Üre$^{4}$%
\thanks{$^{*}$ Equal Contribution}
\thanks{$^{1}$Department of Computer Engineering, Istanbul Technical University, Istanbul, Türkiye.
        {\tt\small kafadar19@itu.edu.tr}}%
\thanks{$^{2}$Department of Computer Science, University of Stuttgart, Stuttgart, Germany.
        {\tt\small st201002@stud.uni-stuttgart.de,}}%
\thanks{$^{3}$Istanbul Technical University Artificial Intelligence and Data Science Application and Research Center, Istanbul, Türkiye.
        {\tt\small \{sanli21, orakf20, gazi20, komurcu17\}@itu.edu.tr}}%
\thanks{$^{4}$Department of Aeronautics and Astronautics, Stanford University, Stanford, CA 94305, U.S.A.
        {\tt\small ure@stanford.edu}}%
\thanks{ Source code is available at: \href{https://github.com/robust-comm-marl-IROS2026/Value-Aware-Prediction-Under-Communication-Loss}{https://github.com/robust-comm-marl-IROS2026/Value-Aware-Prediction-Under-Communication-Loss}}
\thanks{This work was supported in part by the Scientific and Technological Research Council of Türkiye (TÜBİTAK) through the 2210-A Scholarship Program awarded to K. D. Kafadar. \quad {\scriptsize \copyright~2026 IEEE.  Personal use of this material is permitted.  Permission from IEEE must be obtained for all other uses, in any current or future media, including reprinting/republishing this material for advertising or promotional purposes, creating new collective works, for resale or redistribution to servers or lists, or reuse of any copyrighted component of this work in other works.}}
}

\begin{document}
\maketitle
\begin{abstract}
Robust multi-agent coordination relies heavily on inter-agent communication, which is frequently disrupted by physical and environmental constraints in real-world deployments. To maintain operation during these intermittent communication failures, agents can employ internal prediction models to estimate missing shared state information. However, predictors trained with standard reconstruction objectives treat all transitions equally. In a Reinforcement Learning context, this forces the model to waste capacity learning stochastic exploration noise and the outdated dynamics of suboptimal policies. In this paper, we propose a value-aware extension of Multi-Agent Observation Sharing under Communication Dropout (MARO) to patch communication gaps; we refer to this method as \emph{Value-Aware MARO}. By dynamically weighting the predictor's loss function using advantage estimates derived from the underlying actor-critic architecture, our objective explicitly couples the predictor's learning process to the policy's evolution. This formulation focuses the model's capacity on the intentional, high-return dynamics actively reinforced by the agents. We evaluate our framework on several tasks within the Multi-Agent Particle Environment under varying communication reliability levels. Experimental results demonstrate that our approach maintains performance under declining communication reliability, particularly below 40\%. While our method performs comparably in tasks where the baseline already maintains high coordination, our value-aware weighting effectively prevents the performance collapse observed in the standard predictor during high-attrition scenarios. In these environments, our method achieves an average improvement in mean returns of more than 20\% and reduces performance variance by a mean of 64.7\% compared to the standard unweighted baseline.
\end{abstract}

\section{Introduction}



The successful deployment of Multi-Agent Reinforcement Learning (MARL) in cooperative tasks is primarily dependent on robust inter-agent coordination. Since individual agents typically operate under partial observability, they historically rely on persistent communication to share local sensory data and approximate a cohesive global state \cite{das2019tarmac, foerster2016learning}.

Although Reinforcement Learning (RL) provides a powerful, data-driven paradigm for sequential decision-making in uncertain environments \cite{schulman2017proximal}, the assumption of perfect connectivity is unrealistic in practical systems such as autonomous swarms, where signal obstruction or bandwidth constraints lead to intermittent connectivity and risk mission failure \cite{kim2019message, zhang2024multiagent, fu2025multi}. Therefore, a resilient control strategy must handle communication losses gracefully to avoid catastrophic performance degradation \cite{bloom2023decentralized, santos2025centralized, zhang2025pagnet}, particularly in aerial robotics, where RL has already demonstrated effectiveness in managing complex kinematics and momentum \cite{hwangbo2017control, song2021autonomous}.

To address this challenge, researchers have explored internal state predictors that allow agents to estimate missing shared information during communication gaps. Typically, these predictors are trained using standard supervised learning objectives to model the system dynamics \cite{fu2025multi, bloom2023decentralized, zhang2024multiagent, zhang2025pagnet, santos2025centralized}. However, this methodology is fundamentally misaligned with the agents' learning process. It treats all transitions during RL training with the same importance regardless of their relevance to the evolving policies of the agents. By failing to prioritize intentional dynamics, we argue that standard predictors often waste capacity modeling stochastic exploration noise or behaviors that the RL agents have already learned to discard. Building on Multi-Agent Observation Sharing under Communication Dropout (MARO) \cite{santos2025centralized}, we propose a value-weighted predictor objective that leverages the RL agent's internal success signals to emphasize these critical transitions, ensuring reliable estimates when coordination is most vital. We refer to this method as \emph{Value-Aware MARO}.

Our primary contributions in this work are:
\begin{itemize}
    \item A Value-Aware Predictor with an RL-synchronized training flow that utilizes importance weights to prioritize high-consequence transitions over stochastic environment noise.
    \item A coordination framework that patches communication gaps with value-prioritized estimates, allowing decentralized policies to function even under total signal loss.
    \item Empirical validation against a strong baseline in five tasks, demonstrating significant improvements in robustness under severe communication loss.

\end{itemize}


\section{Related Work}

MARL has achieved remarkable success in complex cooperative settings. To address foundational challenges like non-stationarity and partial observability, the Centralized Training with Decentralized Execution (CTDE) framework has become the standard. CTDE leverages global information during training via centralized critics or value factorization to enable scalable, decentralized execution \cite{lowe2017multi, foerster2018counterfactual, rashid2020monotonic, sunehag2017value}. However, while these methods excel under ideal conditions, sustaining robust coordination under communication uncertainty remains a critical open challenge.

Recent works propose differentiable communication learning mechanisms to enhance coordination. Methods such as gradient-based protocols \cite{foerster2016learning}, attention-based selective communication \cite{jiang2018learning}, and targeted message passing \cite{das2019tarmac} demonstrate that structured information exchange improves cooperative behavior. However, most assume reliable communication during execution. In practical multi-agent systems such as UAV swarms, robotic teams, or distributed sensor networks, communication can be intermittent, delayed, or lost due to environmental or hardware constraints.

A growing body of literature studies MARL under communication constraints. Approaches include stochastic message dropout during training \cite{kim2019message}, recurrent policy networks to maintain memory of past observations \cite{wang2020r}, and adaptive coordination strategies for heterogeneous systems \cite{ure2015online}. Although these methods improve robustness against communication degradation, they generally do not explicitly reconstruct missing global information and often rely on structured or periodic assumptions about communication patterns.

To directly address missing information, several works incorporate predictive modeling. Methods include employing Adaptive Predictive Variational Autoencoders \cite{fu2025multi} and intention-sharing frameworks that encode predicted future trajectories \cite{kim2020communication}. More directly related to communication loss, Centralized Training with Hybrid Execution (CTHE) via Predictive Observation Imputation \cite{santos2025centralized} imputes missing observations during decentralized execution using a predictor trained with centralized information. Collectively, these studies establish the importance of predictive reconstruction mechanisms; however, they primarily rely on standard reconstruction objectives that treat all state transitions with equal importance during training.

Taken together, these studies motivate a next step that not only reconstructs missing information, but also emphasizes robustness where coordination is most critical. Our approach follows this direction by incorporating value awareness into predictive imputation.

\section{Background}
\label{sec:background}

\subsection{Problem Formulation: H-POMDP}


We model the multi-agent coordination problem under intermittent communication using the Hybrid Partially Observable Markov Decision Process (H-POMDP) framework  introduced by Santos et al. \cite{santos2025centralized}. Operating over discrete time steps $t \in \mathbb{N}$, an H-POMDP extends the standard Decentralized POMDP (Dec-POMDP) via the tuple $([n], \mathcal{S}, \mathcal{A}, \mathcal{P}, r, \gamma, \Omega, \mathcal{O}, \mathcal{C})$. These components govern the $n$ agents, state space $\mathcal{S}$, joint actions $\mathcal{A}$, transition probabilities $\mathcal{P}$, rewards $r$, and discount factor $\gamma \in [0,1)$. The joint observation space is denoted by $\Omega$, with the observation probability function formalized as $\mathcal{O}(o|s,a)$. Crucially, at any step $t$, the joint observation $o_t \in \Omega$ is the concatenation of all individual local observations, $o_t = \{o_t^1, ..., o_t^n\}$.


The defining feature of this formulation is the $n \times n$ communication matrix $\mathcal{C}$. At any time step $t$, the matrix entry $[\mathcal{C}]_{i,j} = p_{i,j}$ denotes the probability of agent $i$ successfully receiving agent $j$'s local observation $o_t^j$. In a physical robotics context, this localized payload $o_t^j$ typically contains agent $j$'s immediate sensor readouts, such as its spatial coordinates, velocities, or detected target landmarks, as in our experimental tasks. By varying $\mathcal{C}$, this framework naturally interpolates between fully decentralized ($\mathcal{C} = I$) and fully centralized execution.

In our target-search scenarios, agents face an unknown communication matrix $C$ sampled from a distribution $\mu$ during execution. We specifically focus on a probabilistic failure model where shared connections drop with probability $1-p$. Consequently, an agent $i$ either receives the joint observation $o_t$ or falls back to a partial local view.

\subsection{Actor-Critic Architecture in MARL}
CTDE bridges the gap between the necessity of decentralized execution, dictated by the H-POMDP's communication constraints, and sample-efficient learning. Our framework builds upon the Proximal Policy Optimization (PPO) algorithm \cite{schulman2017proximal}, specifically its multi-agent extensions: Independent PPO (IPPO) and Multi-Agent PPO (MAPPO) \cite{yu2022surprising}. 

Both algorithms utilize an actor-critic architecture. The actor networks, $\pi_\theta^i(a_t^i | o_t^i)$, represent decentralized policies mapping local observations to actions. The critic network, $V_\omega(\cdot)$, evaluates the quality of these actions by computing an advantage estimate $\hat{A}_t$, typically via Generalized Advantage Estimation (GAE) \cite{schulman2015high}. The critic differs by algorithm: MAPPO conditions on joint information with a single global network, while IPPO uses each agent's local observation with separate networks. GAE computes the exponentially weighted sum of temporal difference (TD) errors, as defined in \eqref{eq:advantage_eq}:
\begin{equation}
    \hat{A}_t = \sum_{l=0}^{\infty} (\gamma \lambda)^l \delta_{t+l}
    \label{eq:advantage_eq}
\end{equation}

where $\delta_t = r_t + \gamma V_\omega(o_{t+1}) - V_\omega(o_t)$ is the TD error, $\gamma$ is the discount factor, and $\lambda$ is a hyperparameter controlling the bias-variance tradeoff. This advantage quantifies how much better a taken action was compared to the policy's average expectation.

\subsection{Observation Imputation}
To overcome the intermittent communication failures defined by the H-POMDP, our work builds upon the MARO architecture \cite{santos2025centralized}. MARO is designed for hybrid execution in MARL, enabling cooperative agents to operate effectively under arbitrary and dynamic communication levels by employing an autoregressive predictive model that estimates missing information from prior observations.

During the centralized training phase, a Long Short-Term Memory (LSTM)-based transition model is trained to predict next-step observation deltas, denoted as $\Delta o_t$, given the current joint observations $o_t$ and a recurrent history state $h_t$. The parameters of this predictive model, $\phi$, are optimized via supervised learning by minimizing the negative log-likelihood (NLL) of the target next-step deltas across all agents, as formulated in \eqref{eq:maro}:
\begin{equation}
    \mathcal{L}_{\text{MARO}}(\phi) = - \sum_{i=1}^{n} \log p_\phi(\Delta o_t^i | o_t, h_t)
    \label{eq:maro}
\end{equation}

During decentralized execution, each agent $i$ maintains an independent instance of this predictive model to impute any dropped observations from teammates. This preserves a stable, estimated joint-observation state for the policy to act upon. 

While MARO provides a robust architectural foundation for patching missing data, its standard NLL objective treats all observed transitions equally. Section \ref{sec:methodology} details our proposed framework to directly address this fundamental misalignment.

\section{Methodology}
\label{sec:methodology}

\begin{figure*}[t]
    \centering
    \begin{minipage}{0.48\textwidth}
        \centering
        \includegraphics[width=\linewidth]{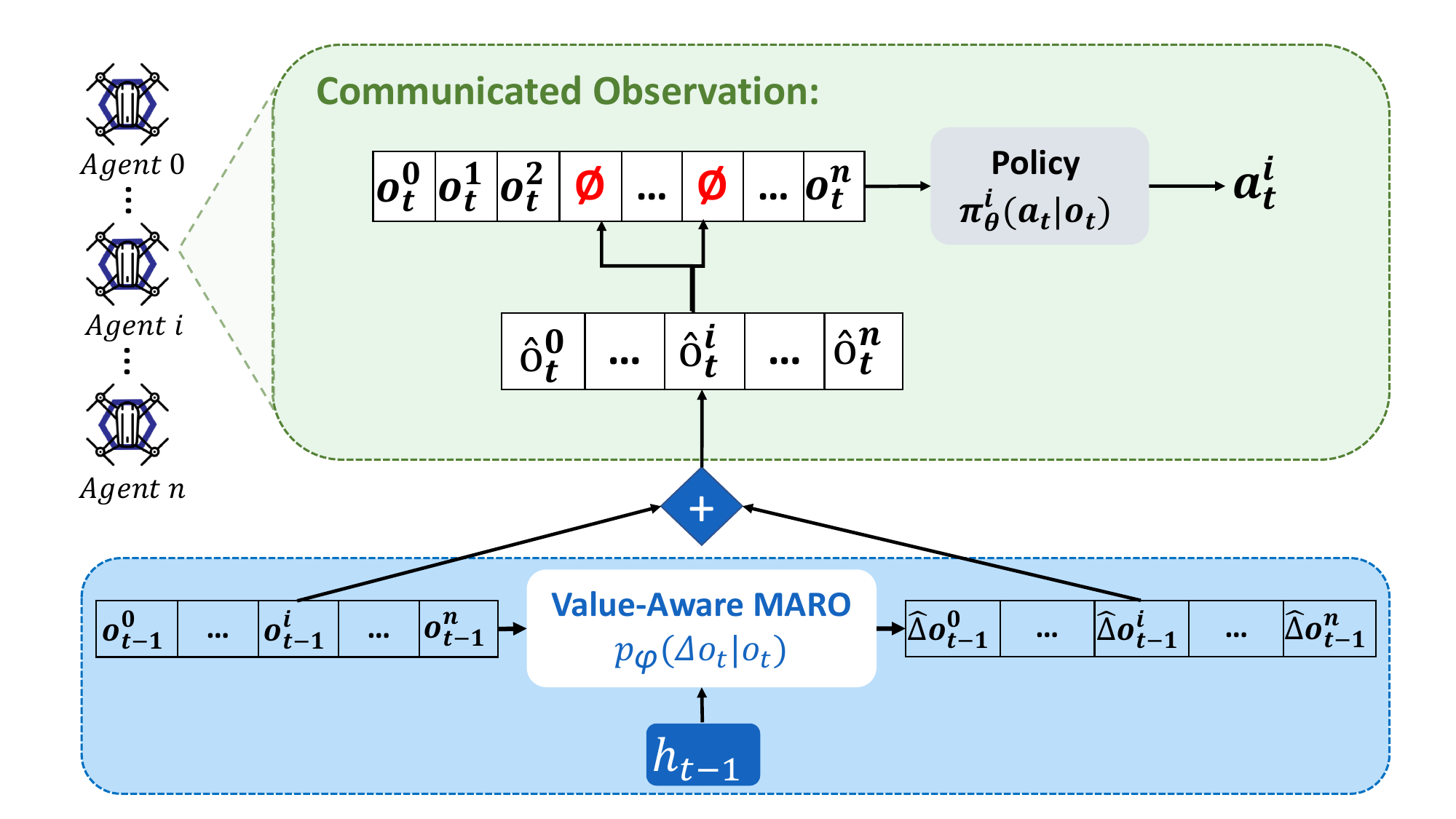}
        \caption{\textbf{MARLxPredictor Integration During Execution:} The predictor intercepts local observations to patch missing shared data during communication failures, providing a reconstructed state to the policy where $o^i_t$ is local observation of agent $i$ at time $t$ and $h_{t-1}$ is the hidden state of the LSTM-based predictor network.}
        \label{fig:system_architecture}
    \end{minipage}
    \hfill
    \begin{minipage}{0.48\textwidth}
        \centering
        \includegraphics[width=\linewidth]{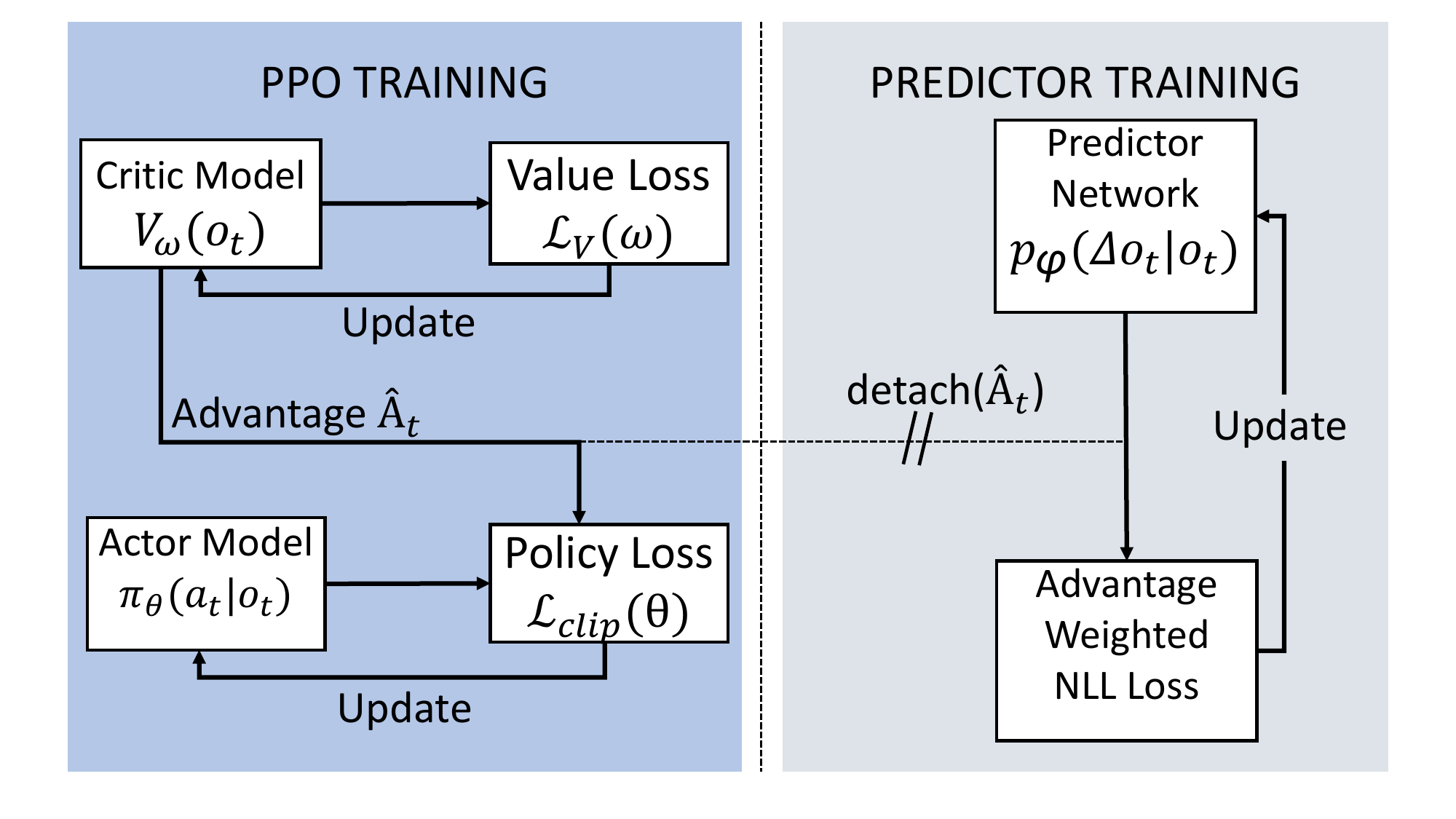}
        \caption{\textbf{Value-Aware Training Pipeline:} The RL training phase generates advantage estimates $\hat{A}_t$, which are detached and used as importance weights for the predictor's Advantage Weighted NLL Loss (formally defined as $J'_{\text{batch}}(\phi)$), ensuring focus on intentional dynamics.}
        \label{fig:training_flow}
    \end{minipage}
\end{figure*}

In this section, we present a control framework designed to sustain multi-agent coordination under varying degrees of communication failure. We adopt a CTDE architecture where agents rely on local sensor data and intermittent shared state information from allies. To bridge the informational gaps caused by signal loss, we integrate a standalone prediction model that imputes the missing parts of the shared information and operates in parallel with the policy network. 

Unlike standard approaches that treat all state dynamics equally, as discussed in previous sections, our framework introduces a value-aware training objective. This mechanism couples the predictor's learning process directly to the policy's evolution, ensuring that the generated state estimates during communication blackouts are optimized to maintain cooperative performance rather than minimizing generic reconstruction error. 

Operating within the H-POMDP framework detailed in Section \ref{sec:background}, our system actively mitigates the performance degradation caused by probabilistic signal drops. We achieve this by predicting the missing shared information, enabling the decentralized policy $\pi_\theta^i(a_t^i | \hat{o}_t)$ to act on an approximated joint observation and preserve swarm coordination despite severe communication loss. We detail the prediction architecture and derive our proposed objective in the remainder of this section.

\subsection{Observation Prediction under Signal Loss}
The employed prediction model, $p_\phi(\Delta o_t | o_t)$, is tasked with estimating the evolution of the environment dynamics. Rather than predicting the next full observation directly, the model learns to predict the change in observations, denoted as $\Delta o_t = o_{t+1} - o_t$. During execution, the predictor intercepts the raw sensory stream before it reaches the policy network and continuously models the environment in parallel, following the system architecture illustrated in Fig.~\ref{fig:system_architecture}.


At each step, it utilizes the available joint observation from the previous timestep, $o_{t-1}$, to estimate the forward transition $\hat{\Delta}o_{t-1} \sim p_\phi (\Delta o_{t-1}| o_{t-1})$, computing an expected current full observation $\hat{o}_{t} \approx o_{t-1} + \hat{\Delta}o_{t-1}$. 

The predictor then evaluates the integrity of the communication link. If the signal is fully intact, the policy receives the ground-truth observation. However, upon detecting signal loss with an agent, the predictor estimates the missing communicated components in the agent's current partial observation. This reconstructed observation is then passed to the policy $\pi_\theta$, which processes it as an imitation of the real observation, allowing the agent to continue its cooperative task without explicit awareness of the underlying communication failure.

\subsection{Value-Aware Predictor Objective}
%


We formalize the Value-Aware objective illustrated in Fig.~\ref{fig:training_flow}, which addresses the equal-weighting limitation of the existing methods discussed in Sec.~\ref{sec:background} by scaling each transition's contribution using the critic's advantage estimate, focusing learning on intentional, high-return dynamics.

In practice, this objective serves as the empirical loss that is minimized during the predictor's optimization step. Formally, our proposed objective for the prediction model over a batch of transitions is defined in \eqref{eq:final-objective}:
\begin{equation}
    J'_{\text{batch}}(\phi) = \mathbb{E}_t \left[ w_t \left( -\log p_\phi(\Delta o_t | o_t) \right) \right]
\label{eq:final-objective}
\end{equation}
where the normalized advantage-based importance weight 
$w_t = \tilde{w}_t \,/\, \mathbb{E}_{t'}[\tilde{w}_{t'}]$ 
with $\tilde{w}_t = \text{ReLU}(1 + \lambda \hat{A}_t)$ 
is derived directly from the critic's advantage estimate 
$\hat{A}_t$, and $\lambda \ge 0$ is a scaling hyperparameter.

To obtain our objective, we derive the predictor's loss evolution with respect to the policy's parameter updates. The objective function of the predictor model for a single transition $(o, \Delta o)$ is defined as the NLL of the observed changes in \eqref{eq:predictor-objective}:
\begin{equation}
    J(\phi) = -\log p_\phi(\Delta o | o)
    \label{eq:predictor-objective}
\end{equation}





where $\Delta o = \mathcal{T}(a, o) - o$ represents the change in observations. Here, $\mathcal{T}$ is the environment's transition function, and the joint action $a$ is sampled from the joint policy $a \sim \pi_\theta(a|o) = \prod_{i=1}^n \pi_\theta^i(a^i|o^i)$. For notational simplicity in the following derivation, we treat the decentralized multi-agent system as a single coordinated entity and derive the objective with respect to the joint policy parameters $\theta$.

Substituting these into \eqref{eq:predictor-objective}, we obtain \eqref{eq:predictor-obj-transition}:
\begin{equation}
    J(\phi) = -\log p_\phi(\mathcal{T}(a, o) - o | o)
    \label{eq:predictor-obj-transition}
\end{equation}

To understand how the predictor's loss changes as the policy parameters $\theta$ evolve, we consider the first-order Taylor expansion of \eqref{eq:predictor-obj-transition} with respect to $\theta$, as formulated in \eqref{eq:taylor-expansion}:
\begin{equation}
    J'(\phi) \approx J(\phi) + \Delta J(\phi) \approx J(\phi) + \nabla_\theta J(\phi)^T \Delta \theta
    \label{eq:taylor-expansion}
\end{equation}

The gradient $\nabla_\theta J(\phi)$ describes how an infinitesimal change in the policy's action selection affects the resulting predictor loss. We derive this relationship in \eqref{eq:grad-link} by using the log-derivative trick:
\begin{align}
    \nabla_\theta J(\phi) &= \nabla_\theta \int \pi_\theta(a | o) [-\log p_\phi(\Delta o | o)] \, da \nonumber \\
    &= \int \pi_\theta(a | o) \nabla_\theta \log \pi_\theta(a | o) [-\log p_\phi(\Delta o | o)] \, da \nonumber \\
    &= \mathbb{E}_{a \sim \pi_\theta} \left[ \nabla_\theta \log \pi_\theta(a | o) \left( -\log p_\phi(\Delta o | o) \right) \right]
    \label{eq:grad-link}
\end{align}

In PPO, the policy parameters are updated to maximize a clipped surrogate objective, which prevents excessively large policy updates. For a given state-action pair $(o, a)$, the objective is defined in \eqref{eq:ppo-clip}:
\begin{equation}
    L^{CLIP}(\theta) = \mathbb{E}_t \left[ \min \left( r_t(\theta) \hat{A}_t, \text{clip}(r_t(\theta), 1 - \epsilon, 1 + \epsilon) \hat{A}_t \right) \right]
    \label{eq:ppo-clip}
\end{equation}
where $r_t(\theta) = \frac{\pi_\theta(a_t | o_t)}{\pi_{\theta_{\text{old}}}(a_t | o_t)}$ is the probability ratio. To determine the update direction $\Delta \theta$, we consider the gradient of this objective. Within the trust region, the gradient with respect to $\theta$ is evaluated in \eqref{eq:ppo-grad}:
\begin{align}
    \nabla_\theta L^{CLIP}(\theta) &\approx \mathbb{E}_t \left[ \frac{\nabla_\theta \pi_\theta(a_t | o_t)}{\pi_{\theta_{\text{old}}}(a_t | o_t)} \hat{A}_t \right] \nonumber \\
    &= \mathbb{E}_t \left[ \frac{\pi_\theta(a_t | o_t)}{\pi_{\theta_{\text{old}}}(a_t | o_t)} \nabla_\theta \log \pi_\theta(a_t | o_t) \hat{A}_t \right]
    \label{eq:ppo-grad}
\end{align}
Evaluating this gradient at $\theta = \theta_{\text{old}}$ simplifies the ratio $r_t(\theta)$ to unity. The resulting parameter update $\Delta \theta$ for a single transition is proportional to this stochastic gradient, as formulated in \eqref{eq:delta-theta}:
\begin{equation}
    \Delta \theta \propto \alpha \hat{A} \nabla_\theta \log \pi_\theta(a | o)
    \label{eq:delta-theta}
\end{equation}
where $\alpha$ is a positive constant. This term represents the direction in the parameter space that most efficiently increases the expected advantage for the current transition.
To motivate our surrogate predictor objective $J'(\phi)$, we substitute the expression for $\nabla_\theta J(\phi)$ from \eqref{eq:grad-link} for a single transition, and the policy parameter update $\Delta \theta$ from \eqref{eq:delta-theta} into the first-order Taylor expansion in \eqref{eq:taylor-expansion}, yielding \eqref{eq:j-prime-expand}:

\begin{align}
    J'(\phi) &\approx J(\phi) + \nabla_\theta J(\phi)^T \Delta \theta \nonumber \\
    &\approx J(\phi) + \left( \nabla_\theta \log \pi_\theta(a | o) [-\log p_\phi(\Delta o | o)] \right)^T \nonumber \\
    &\quad \cdot \left( \alpha \hat{A} \nabla_\theta \log \pi_\theta(a | o) \right) \nonumber \\
    &= J(\phi) + \alpha \hat{A} \|\nabla_\theta \log \pi_\theta(a | o)\|^2 [-\log p_\phi(\Delta o | o)]
    \label{eq:j-prime-expand}
\end{align}
The term $\|\nabla_\theta \log \pi_\theta(a | o)\|^2$ in \eqref{eq:j-prime-expand} represents the squared magnitude of the policy gradient for the given transition. While this magnitude varies per sample, it is a strictly non-negative scaling factor that does not change the direction of the advantage signal. To formulate a practical surrogate objective, we employ a heuristic approximation: we treat the product of the constant $\alpha$ and this varying magnitude as a single constant hyperparameter $\lambda \ge 0$. This greatly simplifies optimization and reduces computational overhead during training.
Substituting $J(\phi) = -\log p_\phi(\Delta o | o)$ back into the expression, we obtain the heuristically weighted loss in \eqref{eq:j-prime-final}:
\begin{align}
    J'(\phi) &\approx [-\log p_\phi(\Delta o | o)] + \lambda \hat{A} [-\log p_\phi(\Delta o | o)] \nonumber \\
    &= (1 + \lambda \hat{A}) [-\log p_\phi(\Delta o | o)]
    \label{eq:j-prime-final}
\end{align}

This formulation effectively weights the predictor's loss based on the advantage $\hat{A}$ calculated by the RL agent. It ensures that the model prioritizes learning from transitions that contribute most significantly to the policy update.

While the previous theoretical motivation focuses on the 
point-wise impact of a single transition, our final surrogate 
objective must operate over mini-batches sampled from a replay 
buffer. To transition to a stable, implementable empirical 
loss, we first define the unnormalized importance weights as 
$\tilde{w}_t = \text{ReLU}(1 + \lambda \hat{A}_t)$. The ReLU operator ensures non-negativity, preventing gradient inversion from highly negative advantages and the resulting destabilization of training. We then normalize these weights across the mini-batch to yield $w_t$, stabilizing the effective update scale as advantage magnitudes fluctuate during training. This results in our final, batch-level surrogate objective function for the prediction model, shown in 
\eqref{eq:final-batch-objective}:
\begin{equation}
    J'_{\text{batch}}(\phi) = \mathbb{E}_t \left[ w_t \left( -\log p_\phi(\Delta o_t | o_t) \right) \right]
\label{eq:final-batch-objective}
\end{equation}

Ultimately, the proposed objective $J'_{\text{batch}}(\phi)$ in \eqref{eq:final-batch-objective} acts as a value-aware surrogate loss. By dynamically weighting the predictor's NLL, the model explicitly filters stochastic exploration noise and prioritizes transitions that align with the current cooperative strategy. This ensures the predictor's capacity is dedicated to modeling the intentional dynamics that generate high returns, naturally tracking shifts in behavior as the policy evolves. This formulation serves as the core predictive objective utilized throughout our experiments.

\section{Experiments}
\label{sec:experiments}

\subsection{Experimental Setup}
\label{subsec:setup}

The performance of Value-Aware MARO is evaluated across five multi-agent coordination tasks. These tasks, derived from the benchmark suite proposed by Santos et al. \cite{santos2025centralized}, are designed as H-POMDPs to emphasize the necessity of inter-agent information sharing. The information exchanged in these tasks, such as ally and target coordinates, mirrors what a physical swarm can share and lose under degraded communication. They utilize the Multi-Agent Particle Environment (MPE) physics engine \cite{terry2021pettingzoo}, incorporating continuous-space kinematics, inertia, and friction to simulate realistic multi-agent interactions where agents try to reach targets and avoid collisions by communicating. Detailed physical dynamics for the environment are provided in Appendix \ref{app:env_dynamics} with illustrations in Fig. \ref{fig:env_illustrations}. The tasks used are:

\begin{itemize}
    \item \textbf{HearSee (HS):} A heterogeneous task where the ``Hear'' agent observes the target landmark but does not know its own coordinates, while the ``See'' agent observes agent coordinates but not the target.
    \item \textbf{SpeakerListener (SL):} A stationary ``Speaker'' observes a target and must guide a mobile ``Listener'' who is blind to the goal. 
    \item \textbf{SimpleSpreadXY-2 \& SimpleSpreadXY-4 (SXY-2 \& SXY-4):} Navigation tasks with 2 or 4 agents where they observe only one axis ($x$ or $y$) and must communicate to resolve the full joint observation. 
    \item \textbf{SpreadBlindfold (SBF):} A three-agent task where local observations are restricted to an agent's own state, requiring communication to localize peers.
\end{itemize}

As established in our H-POMDP formalization (Sec. \ref{sec:background}), we model communication failures probabilistically. We evaluate the system across connectivity levels $p \in \{1.0, 0.8, 0.6, 0.4, 0.2, 0.0\}$, representing transitions from full centralization to total communication blackout.

\subsubsection{Training and Evaluation Protocol}
Following the CTHE paradigm, we utilize MAPPO as the underlying RL algorithm. To isolate the impact of our proposed objective, we compare the baseline MARO against our Value-Aware MARO. Tables \ref{tab:rl_hypers} and \ref{tab:predictor_hypers} in Appendix \ref{app:hyperparams} show details about the configuration for RL and predictive models. Note that all models were trained with a $p=1.0$ sampling scheme for communication to ensure the availability of ground-truth signals for the advantage-weighted objective.

\subsection{Results and Analysis}
\label{sec:results_analysis}

\begin{figure}[!htbp]
    \centering
    \includegraphics[width=1.0\linewidth]{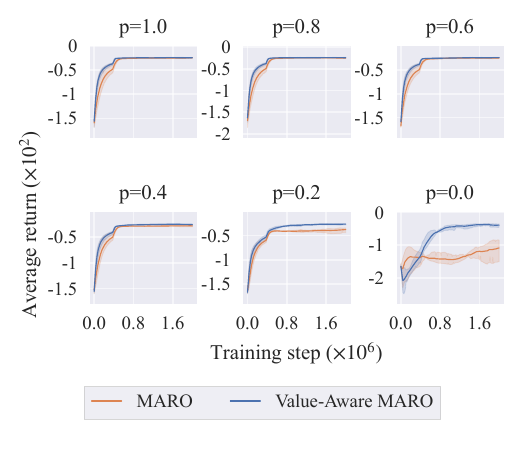}
    \caption{MAPPOxMARO comparisons between performances of evaluations throughout training in the HS environment. We evaluate the average return across various levels of communication constraints ($p \in \{1.0, 0.8, 0.6, 0.4, 0.2, 0.0\}$). The plots demonstrate that our proposed Value-Aware MARO method maintains higher performance compared to the baseline MARO, particularly as communication bandwidth becomes increasingly restricted.}
    \label{fig:BlindDeaf_comp_rew_training_six}
\end{figure}

The training results for the HS task through training are shown in Fig. \ref{fig:BlindDeaf_comp_rew_training_six}. In the extreme case of total communication loss ($p=0.0$), the baseline MARO fails to converge to a coordinated policy, resulting in a significantly lower average return. In contrast, the Value-Aware objective allows the agents to reconstruct the specific observation deltas that correlate with high-advantage states. This focused prediction enables the swarm to maintain an effective estimation of the observations, leading to a higher and more stable reward trajectory throughout the training process. We also provide all training curves for the remaining tasks for MAPPO and also IPPO to show generalization across algorithms in Figs. \ref{fig:appendix_training_grids} and \ref{fig:appendix_ippo_grids} in Appendix \ref{app:extended_results}.

\begin{table}[htbp]
\centering
\captionsetup{labelsep=period, justification=raggedright, singlelinecheck=false, font=small} 
\resizebox{\columnwidth}{!}{%
\begin{tabular}{@{}llccc@{}}
\toprule
\textbf{Environment} & \textbf{Predictor} & \textbf{No Comm ($p=0.0$)} & \textbf{Low Comm ($p=0.2$)} & \textbf{Full Comm ($p=1.0$)} \\ \midrule
{HS} & Baseline & $-102.6 \pm 26.9$ & $-35.9 \pm 4.4$ & $-24.4 \pm 0.7$ \\
 & \textbf{Value-Aware} & $\mathbf{-37.5 \pm 2.9}$ & $\mathbf{-26.3 \pm 1.2}$ & $-24.3 \pm 1.2$ \\ \midrule
{SXY-4} & Baseline & $-1264.4 \pm 19.9$ & $-981.7 \pm 17.6$ & $-834.4 \pm 11.4$ \\
 & \textbf{Value-Aware} & $\mathbf{-1133.6 \pm 9.3}$ & $\mathbf{-937.0 \pm 15.8}$ & $-834.4 \pm 11.4$ \\ \midrule
{SXY-2} & Baseline & $-213.8 \pm 20.6$ & $-171.4 \pm 6.5$ & $-153.8 \pm 0.7$ \\
 & \textbf{Value-Aware} & $\mathbf{-188.9 \pm 2.6}$ & $\mathbf{-163.5 \pm 1.6}$ & $-153.8 \pm 0.7$ \\ \midrule
{SBF} & Baseline & $-422.0 \pm 3.9$ & $-421.5 \pm 5.1$ & $-423.4 \pm 2.5$ \\
 & \textbf{Value-Aware} & $\mathbf{-424.6 \pm 2.9}$ & $\mathbf{-423.5 \pm 4.5}$ & $-423.4 \pm 4.3$ \\ \midrule
{SL} & Baseline & $-27.0 \pm 3.5$ & $-26.6 \pm 1.1$ & $-27.2 \pm 2.2$ \\
 & \textbf{Value-Aware} & $\mathbf{-27.4 \pm 0.9}$ & $\mathbf{-29.0 \pm 0.2}$ & $-27.2 \pm 1.0$ \\ \bottomrule
\end{tabular}
}
\vspace{0.1cm}
\caption{Average evaluation returns after training. Our Value-Aware objective prevents the performance collapse seen in the baseline as communication probability $p$ decreases.}
\label{tab:results_mappo}
\end{table}

The converged performance metrics across all five tasks are summarized in Table \ref{tab:results_mappo}. While the baseline MARO's performance drops drastically as communication levels decrease, our advantage-weighted objective improves it significantly across every task where the baseline experiences performance loss, and also decreases the standard deviation. This indicates that the advantage-based weighting prevents the predictor from being distracted by observation noise during RL training.

\begin{figure}[t]
    \centering
    \includegraphics[width=0.95\linewidth]{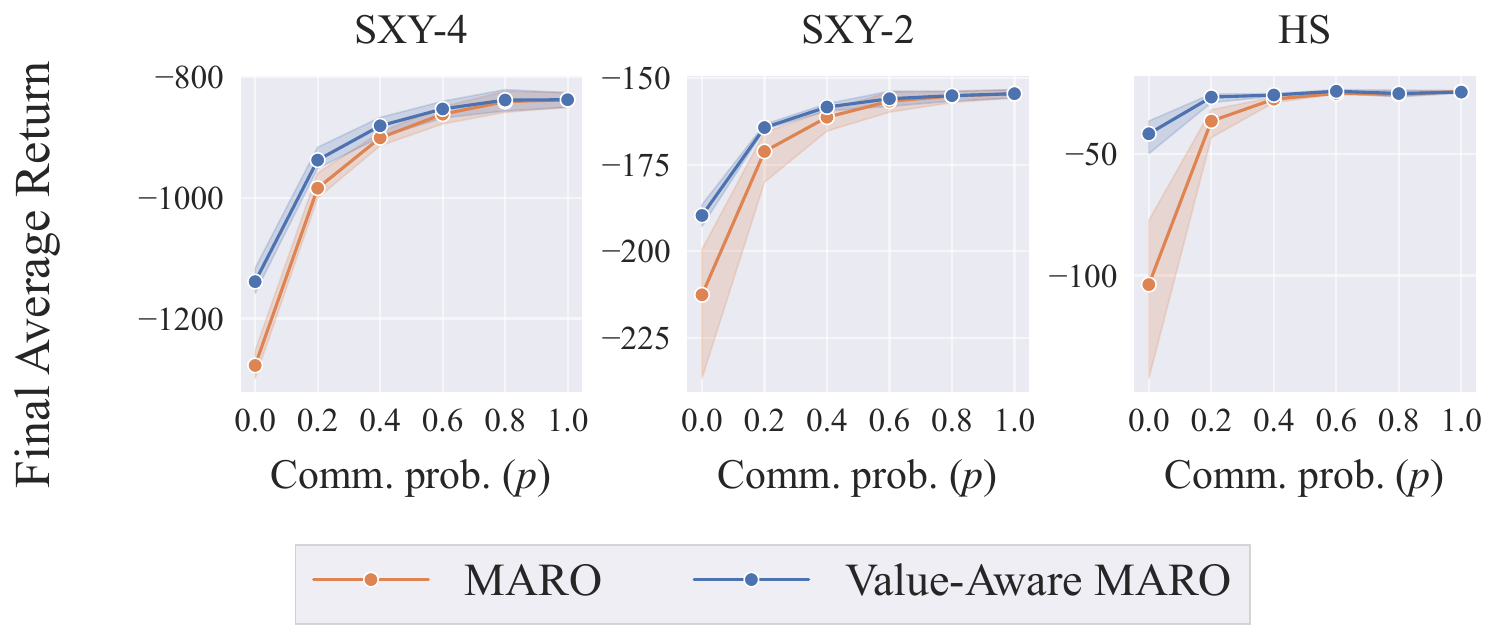}
    \caption{Robustness profile across three coordination tasks: SXY-4, SXY-2, and HS. The plots illustrate the average return as a function of communication probability $p \in [0, 1]$, highlighting the performance of Value-Aware MARO compared to the baseline under varying degrees of connectivity.}
    \label{fig:robustness_profile}
\end{figure}

Finally, Fig. \ref{fig:robustness_profile} illustrates the 
performances of the converged policies under all communication 
levels for the tasks in which the baseline MARO has drastic 
performance drops. While the baseline MARO exhibits a sharp 
performance collapse as $p$ drops below $0.4$, Value-Aware 
MARO maintains a much higher performance. The overlap in 
performance at $p=1.0$ confirms that both methods start with 
the same RL performance, ensuring that the comparisons are 
fair. This shows that any performance gaps at lower 
communication levels are caused solely by the predictors , not by differences in the underlying policy. This 
effect is further amplified as $p$ decreases, where predicted 
observations increasingly dominate the agent's input; 
consequently, the quality of those predictions has a 
compounding effect on policy performance, making the 
value-aware prioritization of high-consequence transitions 
progressively more critical.

\section{Conclusion}
In this study, we introduced an objective that prioritizes the policy's learning signals for prediction models in hybrid execution settings for MARL systems operating under unreliable communication. By weighting the predictor's loss with the RL advantage signal, our framework ensures that models track the intentional dynamics of the cooperative strategy rather than environment noise. Our results across five tasks show that our approach improves performance, particularly in high communication loss scenarios.

The current study evaluates up to four agents on MPE tasks under a probabilistic communication dropout model. Extending the framework to larger teams, temporally correlated or bandwidth-limited communication, and more complex environments remains an open direction. Ultimately, deployment on physical drone swarms is a natural next step toward real-world applicability.

\bibliographystyle{IEEEtran}
\bibliography{IEEEabrv,references}

\appendices

\section{Physical Dynamics of MPE}
\label{app:env_dynamics}

We utilize the MPE \cite{terry2021pettingzoo}, a continuous-space physics engine with the following constraints:

\textbf{1) Kinematics and Inertia:} Agents are circular entities with mass $m = 1.0$ kg and radius $r = 0.05$. State transitions follow semi-implicit Euler integration ($dt = 0.1$s), where actions apply forces ($F = 5.0$ N), which leads the predictor to model second-order temporal dynamics.

\textbf{2) Friction and Damping:} A linear damping factor of $0.25$ simulates air resistance, where agents retain 75\% of their velocity per step. The velocity update is:
\begin{equation}
    v_{t+1} = v_t \times (1 - 0.25) + \frac{F}{m} dt
\end{equation}

\textbf{3) Collision Physics:} MPE uses a soft-repulsion model. When distance $d < r_i + r_j$, a $100$ N contact force is applied via a log-sum-exp formulation for differentiable gradients, and these interactions cause high-frequency velocity changes.

\begin{figure}[htbp]
    \centering
    \subfloat[HS]{\includegraphics[width=0.42\linewidth]{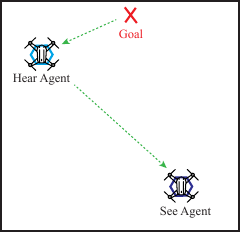}\label{fig:illustration_hearsee}}
    \hfill
    \subfloat[SL]{\includegraphics[width=0.42\linewidth]{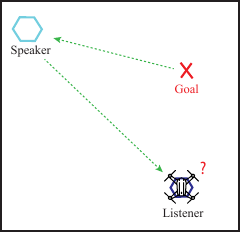}\label{fig:illustration_speaker}}
    
    \vspace{0.2cm} 
    
    \subfloat[SXY]{\includegraphics[width=0.42\linewidth]{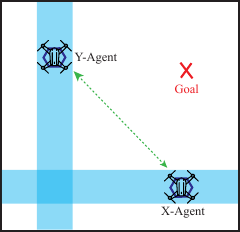}}\label{fig:illustration_XY}
    \hfill
    \subfloat[SBF]{\includegraphics[width=0.42\linewidth]{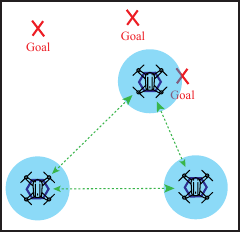}}\label{fig:illustration_blind}
    
    \caption{Representative illustrations of the MPE tasks, demonstrating the specific task objectives and communication-based dependencies of the swarm.}
    \label{fig:env_illustrations}
\end{figure}

Fig. \ref{fig:env_illustrations} illustrates the environments, highlighting their specific semantics and communication-based dependencies.

\section{Experimental Hyperparameters}
\label{app:hyperparams}

For fair comparisons, the RL agents were trained using two primary configurations as shown in Table \ref{tab:rl_hypers}. Setting 1 was utilized for most environments, while Setting 2 was applied to SXY-4 and SBF since it increased the pure RL performance in those scenarios. Table \ref{tab:predictor_hypers} shows the hyperparameters of the predictor models. Notably, the baseline MARO retains the exact optimal settings prescribed in its original implementation. We preserved these original baseline parameters for a fair comparison. For Value-Aware MARO, the predictor batch size is set to match the RL batch size, as the advantage estimates $\hat{A}_t$ are computed per RL minibatch and we keep them structurally aligned rather than using a separate buffer to store the transitions and sample later. A static advantage weight of $\lambda = 1.0$ is maintained across all tasks to demonstrate robustness without environment-specific tuning.

\begin{table}[htbp]
\centering
\caption{Reinforcement Learning Hyperparameters (MAPPO/IPPO)}
\label{tab:rl_hypers}
\resizebox{\columnwidth}{!}{%
\begin{tabular}{@{}lcc@{}}
\toprule
\textbf{Hyperparameter} & \textbf{Setting 1} & \textbf{Setting 2} \\ \midrule
Learning Rate & $1 \times 10^{-4}$ & $3 \times 10^{-4}$ \\
Entropy Coefficient & $0.001$ & $0.01$ \\
Network Type & MLP & GRU \\
Batch Size & 10 & 10 \\
Hidden Dimensions & 256 & 256 \\
Reward Standardization & True & True \\
n-step & 5 & 5 \\ \bottomrule
\end{tabular}
}
\end{table}

\begin{table}[htbp]
\centering
\caption{Predictor Hyperparameters (Value-Aware MARO)}
\label{tab:predictor_hypers}
\footnotesize 
\begin{tabular}{@{}ll@{}}
\toprule
\textbf{Hyperparameter} & \textbf{Value} \\ \midrule
Architecture & LSTM \\
Hidden Dimension & 128 \\
Learning Rate & $1 \times 10^{-3}$ \\
Advantage Weight ($\lambda$) (VA) & 1.0 \\
Gradient Clipping & 1.0 \\
Batch Size (VA / Baseline) & 10 / 32 \\
Buffer Size & 5,000 \\
Training Comm. Prob ($p$) & 1.0 \\ \bottomrule
\end{tabular}
\end{table}

\section{Extended Experimental Results}
\label{app:extended_results}

\begin{figure*}[htbp]
    \centering
    \subfloat[]{\includegraphics[width=0.235\textwidth]{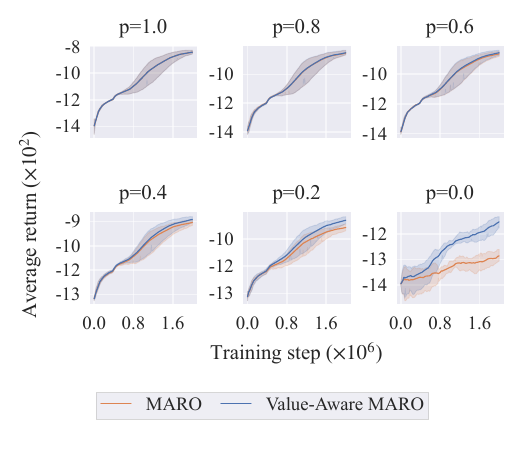}\label{fig:app_xy4}}
    \hfill
    \subfloat[]{\includegraphics[width=0.235\textwidth]{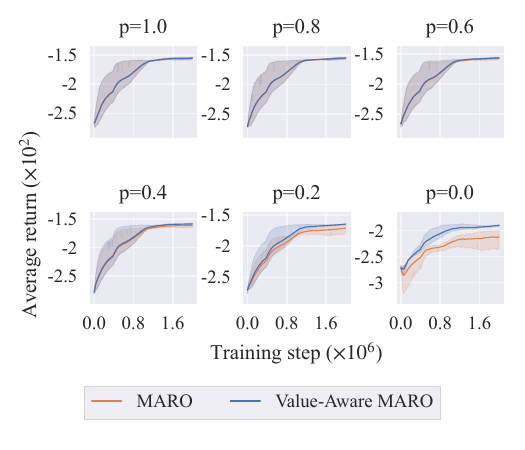}\label{fig:app_xy2}}
    \hfill
    \subfloat[]{\includegraphics[width=0.235\textwidth]{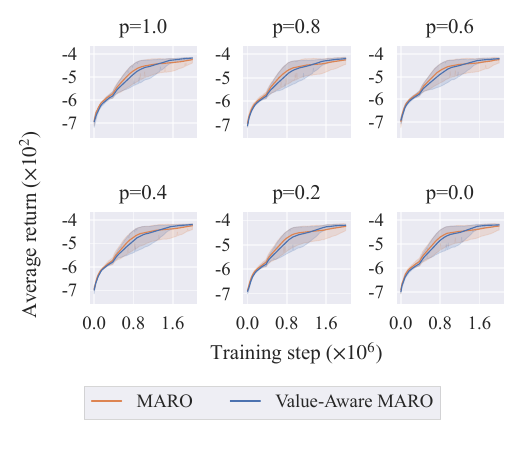}\label{fig:app_blind}}
    \hfill
    \subfloat[]{\includegraphics[width=0.235\textwidth]{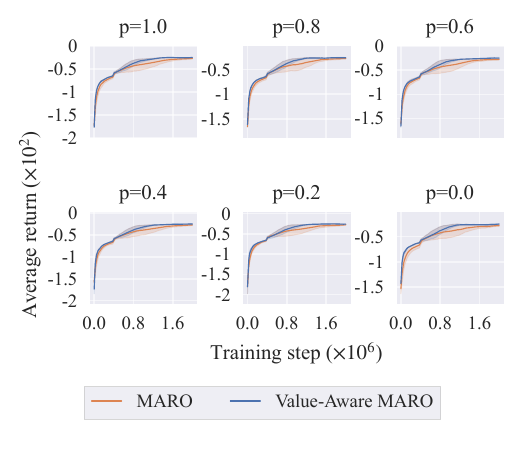}\label{fig:app_speaker}}
    
    \caption{Extended training performance with MAPPO across varying communication levels where $p \in \{1.0, 0.8, 0.6, 0.4, 0.2, 0.0\}$. The tasks shown are: (a) SXY-4, (b) SXY-2, (c) SBF, and (d) SL. The Value-Aware objective (blue) consistently demonstrates superior robustness compared to the baseline MARO (orange).}
    \label{fig:appendix_training_grids}
\end{figure*}

\begin{figure*}[htbp]
    \centering
    \subfloat[SXY-4]{\includegraphics[width=0.31\textwidth]{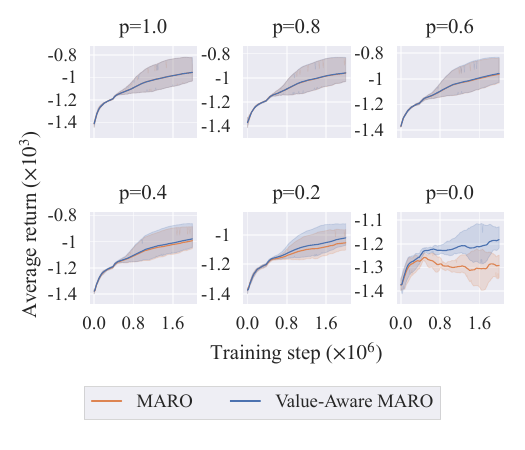}\label{fig:app_ippo_xy4}}
    \hfill
    \subfloat[SXY-2]{\includegraphics[width=0.31\textwidth]{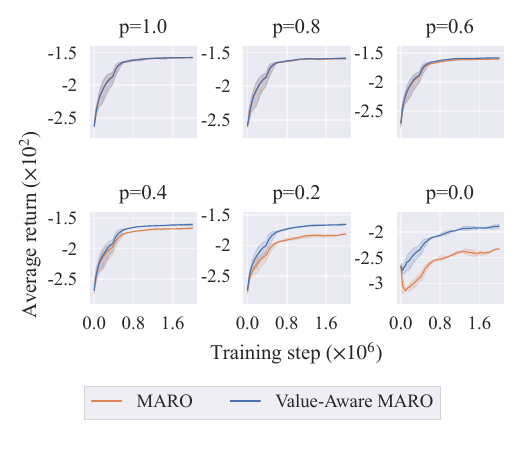}\label{fig:app_ippo_xy2}}
    \hfill
    \subfloat[HS]{\includegraphics[width=0.31\textwidth]{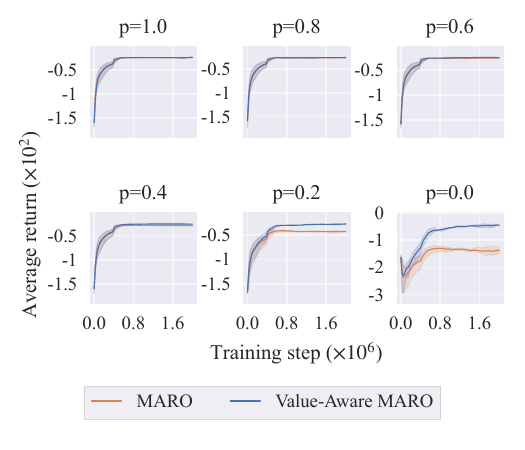}\label{fig:app_ippo_hs}}
    
    \vspace{0.2cm}
    
    \subfloat[SBF]{\includegraphics[width=0.31\textwidth]{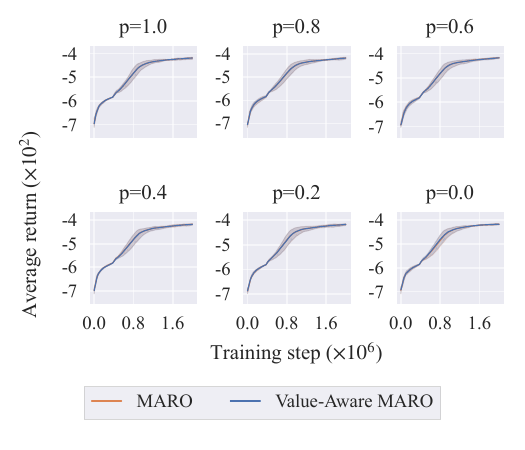}\label{fig:app_ippo_blind}}
    \hspace{0.5cm}
    \subfloat[SL]{\includegraphics[width=0.31\textwidth]{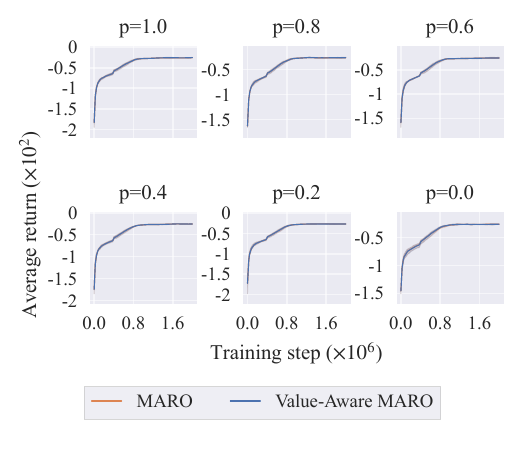}\label{fig:app_ippo_sl}}
    
    \caption{Evaluation curves for IPPO during training of each task are given as comparisons between MARO and Value-Aware MARO to show generalization. The Value-Aware objective remains algorithm-agnostic, providing superior coordination compared to the baseline across both centralized and independent optimization schemes.}
    \label{fig:appendix_ippo_grids}
\end{figure*}

Extended experimental results are given in Figs. \ref{fig:appendix_training_grids} and \ref{fig:appendix_ippo_grids}. For further details about the environments, models and experiments; please check the github repository at: https://github.com/robust-comm-marl-IROS2026/Value-Aware-Prediction-Under-Communication-Loss
 
\end{document}